\documentstyle[preprint,aps,graphicx,epsfig]{revtex}
\begin{document}
\font\mybb=msbm10 at 12pt
\def\bb#1{\hbox{\mybb#1}}
\def\Z {\bb{Z}}
\def\R {\bb{R}}
\def\E {\bb{E}}
\def\unit{\hbox to 3.3pt{\hskip1.3pt \vrule height 7pt width .4pt \hskip.7pt
\vrule height 7.85pt width .4pt \kern-2.4pt
\hrulefill \kern-3pt
\raise 4pt\hbox{\char'40}}}
\def\II{{\unit}}
\def\cM {{\cal{M}}}
\def\half{{\textstyle {1 \over 2}}}

\def    \beq    {\begin{equation}} \def \eeq    {\end{equation}}
\def    \bea    {\begin{eqnarray}} \def \eea    {\end{eqnarray}}
\def    \lf     {\left (} \def  \rt     {\right )}
\def    \a      {\alpha} \def   \lm     {\lambda}
\def    \D      {\Delta} \def   \r      {\rho}
\def    \th     {\theta} \def   \rg     {\sqrt{g}} \def \Slash  {\, /
\! \! \! \!}  \def      \comma  {\; , \; \;} \def       \pl
{\partial} \def         \del    {\nabla}

\preprint{UG-00-08}

\title{Critical fields on the M5--brane and 
noncommutative open strings}
\author{E. Bergshoeff\footnote{e.bergshoeff@phys.rug.nl},
D. S. Berman\footnote{d.berman@phys.rug.nl},
J. P. van der Schaar\footnote{schaar@phys.rug.nl} and
P. Sundell\footnote{p.sundell@phys.rug.nl} }
\address{Institute for Theoretical Physics, University of Groningen, \\
Nijenborgh 4, 9747 AG Groningen, The Netherlands }

\maketitle
\begin{abstract}

The M5--brane is investigated near critical field-strength.
We show that this limit on the M5--brane reduces to the
noncommutative open string limit on the D4--brane. The
reduction on a two--torus leads to both the
noncommutative open string limit and the 
noncommutative Yang--Mills limit on the D3--brane. 
The decoupled noncommutative five--brane is identified
with the strong coupling limit of the noncommutative open string theory on
the D4--brane and S--duality on the noncommutative D3--brane is 
identified with a modular
transformation on the five--brane. 
We argue that the open membrane metric defines a  finite length scale on the
worldvolume of the M5--brane in the decoupling limit. This length scale
can be associated to the effective length scale of an open membrane.

\end{abstract}

\section{Introduction}

Noncommutative geometry has been shown to play a fascinating role in
string theory. The original interest was sparked by the appearance of
noncommutativity on D-branes in the presence of constant background
Neveu--Schwarz two--form potentials $B_{\rm NS}$
\cite{Connes:1998cr,Schwarz:1998qj,sw}.
On the D--brane itself these potentials appear as a two--form adapted field
strength
${\cal{F}}=dA+B_{NS}$. In the magnetic case (i.e.~when ${\cal F}$
has only spatial parts) a spatially noncommutative Yang--Mills
theory (NCYM) on the D--brane can be decoupled from the bulk gravity
\cite{sw}.
The natural, though conceptually challenging generalisation, was to include
field strengths with non-zero electric components
thus inducing spatio-temporal noncommutativity on the D-brane.
This was examined in \cite{GMSS,lennat,Barbon:2000sg,Chen:2000ny,kleb}. 
The somewhat remarkable
result was that by examining the D-brane in a decoupling limit near critical
electric field strength\footnote{The concept of a critical field only works 
for the electric case
as it relies crucially on the Lorentz signature.}
 one naturally constructed a unitary
decoupled spatio-temporally noncommutative open string theory (NCOS).
The thermodynamics of such theories was examined from the dual
supergravity point of view in \cite{troels}.

The crucial property of the NCOS limit described in \cite{GMSS,lennat} is
to keep fixed both the open string two--point function and the effective
open
string coupling constant

\beq
<X^A X^B>=  2 \pi
\alpha^\prime G^{AB} + \Theta^{AB} =
2 \pi \alpha^\prime \left( { 1 \over {g + 2 \pi \alpha^\prime
\cal{F}}}\right)
^{AB} \, ,
\label{ncos2pt}
\eeq
\beq
G_{\rm 0} = g_s { {det^{1/2}(g + 2 \pi \alpha^\prime \cal{F})} \over {
det^{1/2}(g) } }\, ,
\label{ncosg}
\eeq

where $G^{AB}$ is the symmetric part and $\Theta^{AB}$ the antisymmetric
part of the two--point function. 
In this limit the leading divergent parts of $\cal{F}$ cancel
the contribution from  $g_{\mu
\nu}/\alpha'$,
leaving the two--point function and coupling governed by the
finite subleading terms\footnote{Here and in the rest of this paper
we will ignore factors of $2\pi$ etc.}.
As discussed in \cite{GMSS}, this NCOS limit introduces a fixed metric
$G^{AB}_{\rm OS}$ and a 
new effective length scale $\alpha^\prime_{\rm eff}$ defined
as follows:

\beq
\alpha^\prime G^{AB} = \alpha^\prime_{\rm eff} G^{AB}_{\rm OS}\, .
\label{gos}
\eeq

In this paper we will consider a
decoupling limit in the M5--brane analogous to the noncommutative
open string limit and examine its properties through an open membrane
probe.
Even though the analog of the string two--point function is not
available for the open membrane, it was conjectured in \cite{us} that
the decoupled five--brane theory should be formulated in terms
of a so called {\it open membrane
metric} with properties analogous to those of the
open string metric.
Thus we demand that the six--dimensional proper
lengths measured by the open membrane
metric are fixed in units of the 11D Planck length $\ell_p$. 
This defines a finite effective
length scale of the decoupled five--brane theory 
that we shall denote $\ell_g$.
This is suggestive of an open membrane  theory underlying the
decoupled spatio-temporal noncommutative M5--brane.

Evidence for the decoupled non--commutative five--brane
theory can be obtained by comparing various reductions of the
five--brane limit to NCOS and NCYM limits in string theory. 
Conversely, this
provides a direct interpretation of the strong coupling behavior of the
decoupled NCOS and NCYM theories.

The structure of the paper is as follows. We begin in Section II
by describing the decoupling limit on the M5--brane. 
In section III we show that this
limit reduces to the NCOS limit on the
D4--brane. In Section IV we show that the single limit on the M5--brane
reduces to both the NCOS and NCYM limits on the D3--brane.
These two limits are related by a modular transformation
on the two--torus. Here the role of the open membrane metric is shown to
play a crucial role.
We end with some conclusions and discussion.

\section{The M5--brane and the decoupling limit}

The five--brane may be effectively described by a six--dimensional
self-dual two form field theory (this is neglecting the superpartners
in the (2,0) supermultiplet). The adapted field strength is

\beq
{\cal H}= db+C \ ,\label{calh}
\eeq

where $C$ is the pull--back to the five--brane of the three--form potential in
eleven--dimensional supergravity and $b$ is the two form potential on the
five--brane worldvolume. The self--duality condition provides a
nonlinear algebraic constraint involving the components of the field
strength and the induced metric $g_{\mu\nu}$ on the brane as follows
\cite{Howe:1997mx}:

\beq
{\sqrt{-\det g}\over
6}\epsilon_{\mu\nu\rho\sigma\lambda\tau} {\cal
H}^{\sigma\lambda\tau}={1+K\over 2}(G^{-1})_{\mu}{}^{\lambda} {\cal
H}_{\nu\rho\lambda}\ , \label{nlsd}
\eeq

where $\epsilon^{012345}=1$ and the scalar $K$ and the tensor $G_{\mu\nu}$
are
given by

\begin{eqnarray}
\label{k}
K &=&\sqrt{1+{\ell_p^6 \over 24}{\cal{H}}^2}\, ,\\ &&\cr G_{\mu\nu} &=&
{1+K\over 2K}\left(g_{\mu\nu}+{\ell_p^6\over 4} {\cal
H}^2_{\mu\nu}\right)\ ,
\label{om}
\end{eqnarray}

where $\ell_p$ is the 11D Planck scale.

The relation (\ref{nlsd}) involves the metric, the field
strength components and the plank length $\ell_p$. As we want to
carry out a scaling in these quanities we must make sure that any scaling
obeys
the above relation (\ref{nlsd}). This is amply discussed in \cite{sw,us}.
To facilitate this we introduce a
parametrisation of constant flux solutions to (\ref{nlsd}) as follows:

\beq
{\cal H}_{\mu\nu\rho}= {h\over \sqrt{1+\ell_p^6 h^2}}
\epsilon_{\alpha\beta\gamma} v^{\alpha}_{\mu} v^{\beta}_{\nu}
v^{\gamma}_{\rho} +h\, \epsilon_{abc} u^{a}_{\mu} u^{b}_{\nu}
u^{c}_{\rho}\ , 
\label{hsol}
\eeq
\beq G_{\mu\nu}=
{\left( 1+\sqrt{1+h^2\ell_p^6}\right)^2 \over 4}
\left( {1\over 1+h^2\ell_p^6} \eta_{\alpha\beta}v^{\alpha}_{\mu}v^{\beta}_{\nu}
+\delta_{ab}u^{a}_{\mu}u^{b}_{\nu}\right)\ . 
\label{openm}
\eeq

Here $h$ is a real field of dimension (mass)$^3$ and $(v_\mu^\alpha,
u_{\mu}^{a})$, $\alpha=0,1,2$, $a=3,4,5$, are sechsbein fields in the
nine--dimensional coset $SO(5,1)/SO(2,1)\times SO(3)$ satisfying

\beq
g^{\mu\nu}v_{\mu}^{\alpha}v_{\nu}^{\beta}=\eta^{\alpha\beta}\ ,\quad
g^{\mu\nu}u_{\mu}^{a}v_{\nu}^{\beta}=0\ ,\quad
g^{\mu\nu}u_{\mu}^{a}u_{\nu}^{b} = \delta^{ab}\ , \nonumber
\eeq

\beq
g_{\mu\nu} = \eta_{\alpha\beta}v_\mu^\alpha v_\nu^\beta + \delta_{ab}
u_\mu^{a}u_\nu^{b}\, . \label{sol}
\eeq

A derivation of this parametrisation is given in \cite{us}.

The relation between the tensor $G_{\mu \nu}$ and the open membrane metric
for the
five--brane in analogy with the open string metric that occurs on D-branes
was discussed in \cite{us}. It should be noted that the overall conformal
scale of the open membrane metric was not determined.
In this paper such an overall scale will play a role. We therefore define
the open membrane metric as follows

\begin{equation}
\hat{G}_{\mu\nu}\equiv \phi(x)
\left(g_{\mu\nu}+{\ell_p^6\over 4} {\cal H}^2_{\mu\nu}\right)\ ,
\label{ommhat}
\end{equation}
where the function $\phi(x)\neq 0$ and $x$ is given by the dimensionless
combination $x = \ell_p^6{\cal H}^2$. Using the parametrisation
(\ref{hsol}), this metric can be written as

\beq \hat{G}_{\mu\nu}=(1+{1\over 2}h^2\ell_p^6)\, \phi(h^2\ell_p^6) \,  
\left({1\over 1+h^2\ell_p^6} \eta_{\alpha\beta}v^{\alpha}_{\mu}v^{\beta}_{\nu}
+\delta_{ab}u^{a}_{\mu}u^{b}_{\nu} \right)\ . 
\label{openmc}
\eeq
Below we shall determine the asymptotic
behavior of $\phi(x)$ for large $x$ from the requirements of the decoupling
limit.

We now proceed with the definition of the decoupling limit.
The properties that we demand for the decoupling limit we wish to take are
as follows:

\begin{enumerate}
\item[i)] The Planck length $\ell_p\rightarrow 0$, so that the gravitational
interactions can be decoupled.
\item[ii)] The proper six--dimensional lengths $ds^2(\hat{G})$ of the open
membrane metric are fixed in eleven--dimensional Planck units in the limit, 
i.e.~$\ell_p^{-2}ds^2(\hat{G})$ is fixed, so that the limit describes a genuine
six--dimensional theory with a finite length scale $\ell_g$.
\item[iii)] The electric components contain a divergent piece and a constant
piece, in analogy with the limit discussed in \cite{GMSS} for open
strings.
\end{enumerate}

The first condition we satisfy by scaling $\ell_p\sim\epsilon^{1\over 3}$
($\epsilon\rightarrow 0$). 
In order to satisfy the second and third condition
we impose that $h\ell_p^3$
diverges\footnote{This is in contrast with the limit in \cite{us} where
$h\ell_p^3 $ does not diverge.}.

We are therefore led to consider the
following limit:

\begin{eqnarray}
g_{\alpha \beta} \sim \epsilon^0 \, \Rightarrow v \sim \epsilon^{0} ~~ &;&
\quad g_{ab} \sim \epsilon^1 \Rightarrow u \sim \epsilon^{1\over 2} \, ,
\nonumber \\
\ell_p  \sim  \epsilon^{1 \over 3} \, , \quad  h &\sim& \epsilon^{-{3 \over
2}} \, , \quad \epsilon \rightarrow 0 \, .
\label{tnc5lim}
\end{eqnarray}

such that the components of $\cal{H}$ given by (\ref{hsol}) behave
as follows:

\begin{eqnarray}
{\cal{H}}_{012} &\sim&  \ell_p^{-3}(1 - {1 \over 2} \ell_p^{-6} h^{-2})
\sim\epsilon^{-1} + \epsilon^{0} \, , \nonumber \\
{\cal{H}}_{345} &\sim& h u^3\sim \epsilon^{0} \, .
\label{hcomp}
\end{eqnarray}

The physics on the five--brane in the decoupling limit is uniquely defined
by the two fixed noncommutativity parameters of dimension [length]$^2$
constructed from the the finite parts of
${\cal H}_{012}$ and ${\cal H}_{345}$ as follows:

\begin{equation}
\Theta_{\rm T} \equiv (h^2 \, \ell_p^9)^{2/3} \quad , \quad \Theta_{\rm S}
\equiv (h \, u^3)^{-2/3} \, ,
\label{fixedp}
\end{equation}

where we have set $v=1$. 
In order to satisfy requirement (ii) we demand in analogy with
[\ref{gos}]

\beq
\ell_p^2 \, (\hat{G}^{-1})^{\mu \nu} \equiv
\ell_g^2\, G^{\mu\nu}_{\rm OM}\ {\rm is\ fixed}\, .
\eeq
This allows us to fix the conformal factor as follows:

\begin{equation}
\phi(x)\sim x^{-{2\over 3}}\quad{\rm as}\ \ \ x\rightarrow \infty\ .
\label{sds}
\end{equation}
With the conformal factor now fixed we find

\beq
\ell_p^2 \, (\hat{G}^{-1})^{\mu \nu} =
(\Theta_{\rm T} \eta^{\alpha\beta}\oplus \Theta_{\rm S}\delta^{ab})
\equiv \ell_g^2\, G^{\mu\nu}_{\rm OM}\ .
\eeq

This defines a noncommutative M5--brane length scale
$\ell_g$, a fixed metric $G^{\mu\nu}_{\rm OM}$
 and a dimensionless parameter $\lambda$ as follows

\begin{equation}
\ell_g \equiv \sqrt{\Theta_{\rm T}}\ ,\quad \lambda \equiv {\Theta_{\rm S}
\over \Theta_{\rm T}}\, ,\quad
G^{\mu\nu}_{\rm OM} = (\eta^{\alpha\beta}\oplus \lambda\delta^{ab})
\ .
\end{equation}

\section{The NCOS limit on the D4--brane}

In this section we show that the decoupling limit (\ref{tnc5lim}) on the
M--theory five--brane reduces to the NCOS limit
on the D4--brane. This provides an interpretation
of the spatio-temporal noncommutative
five--brane as the strong coupling dual of the NCOS on the D4--brane.

In order to show this we wrap the five--brane, for finite $\epsilon$,
on a circle of fixed radius $R$ in the direction $x^2$ and identify

\begin{equation}
x^2=X^{11} \sim X^{11}+R\, ,
\qquad {\cal F}_{AB}=R {\cal H}_{AB2} , \qquad A,B=0,1,3,4,5 \, .
\end{equation}

Clearly this means that only ${\cal F}_{01}$ is nonzero on the D4--brane.
We also use the following standard relations between M--theory and
IIA string theory parameters:

\begin{equation}
g_s= \left( { R \over \ell_p } \right)^{3 \over 2} \, , \qquad
\alpha^{\prime}=
{ \ell_p^3\over R} \, .
\end{equation}

The scaling of the metric components in $D=11$ induces the same
scaling for the ten--dimensional metric components and together with
the requirement of fixed radius $R$ we find the following limit on the
D4--brane
(we reset our conventions such that $\alpha,\beta=0,1$)

\begin{eqnarray}
g_{\alpha \beta} \sim \epsilon^0 \, , \quad g_{ab} &\sim& \epsilon^1  \, ,
\quad
{\cal F}_{01} \sim \epsilon^{-1} + \epsilon^0\, , \nonumber \\
\alpha^{\prime}  \sim  \epsilon^1 \, , \quad  g_s &\sim& \epsilon^{-{1 \over
2}} \, , \quad \epsilon \rightarrow 0 \, .
\end{eqnarray}

As a result we find that length scales on the D4--brane,
as measured by the open string metric $G^{AB}$, are kept fixed in the
limit. This also holds for the noncommutativity
parameters $\Theta^{AB}$ appearing in the two--point function
(\ref{ncos2pt}) and the open string coupling $G_{\rm O}$
given by (\ref{ncosg}). We identify the
NCOS limit on the D4--brane with electric field strength ${\cal F}_{01}=
{\cal F}_c-{1\over 2}\theta^{-1}$, where the diverging critical electric
field
${\cal F}_c$ and the fixed noncommutativity parameter $\theta$ are given by

\beq
{\cal F}_c=R\ell_p^{-3}\ ,\qquad \theta={h^2\ell_p^9\over R}\ .
\eeq

Hence, using 
(\ref{hcomp}) and (\ref{fixedp}), we can write fixed D4--brane
quantities in terms of the fixed five--brane data $\Theta_{\rm T}$,
$\Theta_{\rm S}$
and $R$, or equivalently $\ell_g$, $\lambda$ and $R$:

\beq
\alpha^{\prime}\, G^{AB} = ({\Theta_{\rm T}^{3\over 2}\over R}\eta^{\alpha
\beta}
\oplus {\Theta_{\rm T}^{1\over 2}\Theta_{\rm S}\over R}\delta^{ab})
={\ell^3_g\over R}(\eta^{\alpha\beta}\oplus \lambda\delta^{ab}) =
{\ell_g^3\over R} G^{AB}_{\rm OM}\, ,
\eeq
\beq
\theta^{AB}=({\Theta_{T}^{3\over 2}\over R}\epsilon^{\alpha\beta}\oplus 0) =
{\ell_g^3\over R}(\epsilon^{\alpha\beta} \oplus 0)\ ,
\eeq
\begin{equation}
G_{\rm O}=\left( {R \over \ell_g} \right)^{3 \over 2} \, .
\end{equation}

Therefore the NCOS limit on the D4--brane has an effective open string scale
$\alpha^\prime_{\rm eff}$ and noncommutativity parameter $\theta$ given by

\begin{equation}
\alpha^\prime_{\rm eff} \equiv \theta = {\ell_g^3 \over R}\, ,\quad
G_{\rm OM}^{AB} = G^{AB}_{\rm OS} \, .
\end{equation}

Thus we find the following relations between open string moduli and
M--theory open membrane moduli:

\begin{eqnarray}
R&=&G_O \sqrt{\alpha^\prime_{\rm eff}} \\
l_g&=& G_O^{1 \over 3} \sqrt{\alpha^\prime_{\rm eff}}\, .
\end{eqnarray}

\begin{figure}[h]
\begin{center} 
\includegraphics[angle=0, width=140mm]{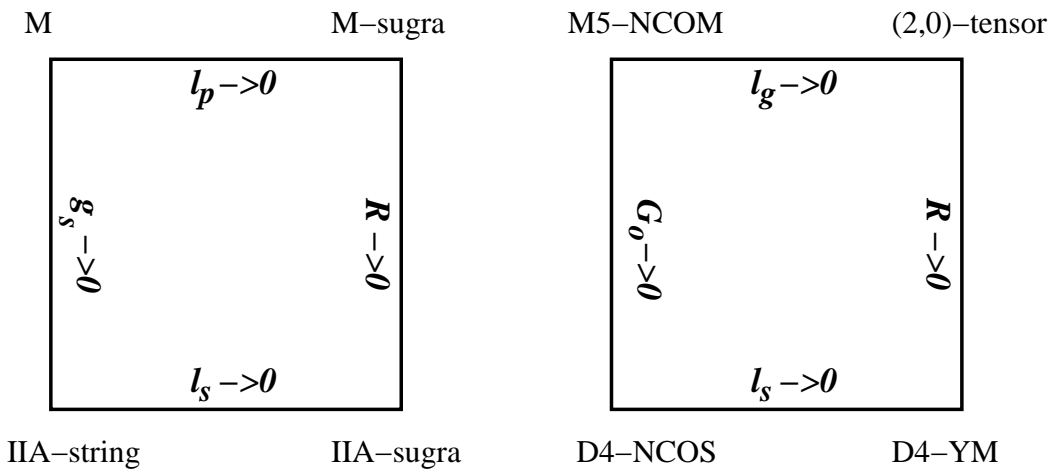} 
\end{center}
\caption{\small The square on the left
relates M-theory and IIA-superstring theory and their low energy limits.
Similarly, the square on the right displays the relations between the
D4-brane noncommutative open string theory (NCOS)  and its M5-brane
noncommutative open membrane (NCOM) origin, and their low energy limits.}
\label{M5qg} 
\end{figure}

These are formally equivalent to the standard relations between
the moduli of M-theory and IIA superstring theory provided that
we give $\ell_g$ a six--dimensional interpretation analogous to
that of the eleven--dimensional Planck scale $\ell_p$ in M--theory. 
This suggests
that the
NCOS theory on the D4--brane generates an extra dimension when we
increase the open string coupling and in the limit $R\rightarrow \infty$ we
end up
with a noncommutative (in all directions!) six--dimensional theory governed
by the
scale $\ell_g$, as displayed in Figure \ref{M5qg}.
Note that $\alpha^\prime_{\rm eff}= \theta$ implies that a field theory
limit
taking $\alpha^\prime_{\rm eff} \rightarrow 0$ will at the same time
also result in vanishing spatio-temporal noncommutativity.

\section{The noncommutative limits on the D3--brane}
   
Next we carry out the double--dimensional reduction of the
M--theory five--brane on a fixed two--torus in order to compare with
the limits given in \cite{GMSS} for the D3--brane
directly reduced on a circle. We drop all Kaluza--Klein modes and identify
the wrapped five--brane with the directly reduced D3--brane in nine
dimensions. This gives
the following relations between M--theory five brane and IIB
three brane 
quantities
\cite{schwarz,aspinwall,Berman:1998va}

\beq
(x^2,x^5)=(X^{11},X^9) \sim (X^{11}+R_2, X^9+R_5) \, ,
\eeq
\begin{eqnarray}
{\cal F}_{AB} = R_2 {\cal H}_{AB 2}  \, , \qquad { g^{(E)}_{AB} \over
\alpha^\prime}
= { \sqrt{R_2 R_5 u} \over \ell_p^3 } g_{AB}\, ,\quad
 \quad A,B=0,1,3,4\, ,
\end{eqnarray}
\begin{equation}
g_s =  { R_2
\over  R_5 u }  \, , \qquad {R_B^{(S)}\over\sqrt{\alpha'}} =
{\ell_p^{3\over 2} \over R_2^{1\over 2} R_5 u} \, .
\end{equation}

where the two--torus coordinate periodicities
$R_2$ and $R_5$ are fixed quantities (we set the
real part of the complex structure equal to zero) 
and where $u$ is the scale of the induced dreibein on
the five--brane in the $3,4,5$ directions. The quantity
$R_B^{(S)}$ is the IIB radius in string frame and $g^{(E)}_{MN}$ is the
IIB Einstein metric which is related to the IIB string metric by
$g^{(E)}_{MN}=g_s^{-{1\over 2}}g^{(S)}_{MN}$. 

The S-dual description of the D3--brane, giving a magnetic field strength,
can be obtained by performing a modular transformation on the two--torus,
which gives the following relations between the quantities in M--theory and
the S-dual picture:

\begin{eqnarray}
\tilde{{\cal F}}_{AB} = R_5 {\cal H}_{AB 5}  \, , \qquad
\tilde{g}_s ={ R_5 u \over { R_2 }}
\, , \qquad  {\tilde{g}^{(E)}_{AB} \over \tilde{\alpha}^\prime} =
{g_{AB}^{(E)} \over \alpha^\prime} \, , \quad {\tilde{R}_B^{(S)} \over
\sqrt{\tilde{\alpha}^\prime} }
 = g_s^{- {1\over 2}} {R_B^{(S)} \over \sqrt{\alpha^\prime}}  \,
,\label{Sdual}
\end{eqnarray}

where the tilde denotes quantities in the the IIB S-dual picture.

Inserting the scaling limit for the M5--brane (\ref{tnc5lim}) in the
first set of relations above we obtain
the
following scaling limit for the D3--brane:

\begin{eqnarray}
{\cal F}_{01} &\sim& \epsilon^{-1} + \epsilon^0 \, , \qquad
{\cal F}_{34} = 0\, , \nonumber\\
{g^{(S)}_{AB} \over \alpha^\prime} &\sim&
{\rm diag}(\epsilon^{-1},\epsilon^{-1},\epsilon^{0},\epsilon^{0})
\, , \qquad g_s \sim \epsilon^{-{1\over 2}}\ .\label{ncos}
\end{eqnarray}

We identify (\ref{ncos}) as the open string limit where
the critical field strength and the noncommutativity parameter $\theta_{\rm 
NCOS}$
are given in terms of M--theory quantities by

\beq
{\cal F}_c=R_2 \ell_p^{-3} \, , \qquad \theta_{\rm NCOS}= 
{ h^2 \ell_p^9 \over
R_2} \ .
\eeq

Identifying the finite quantities on the D3--brane with the
finite quantities on the wrapped M five--brane we obtain the following
relations (we reset our conventions such that $\alpha=0,1$ and $a=3,4$):

\beq
\alpha'G^{AB}=({\Theta_{\rm T}^{3\over 2}\over R_2}\eta^{\alpha\beta} \oplus
{\Theta_{\rm T}^{1\over 2}\Theta_{\rm S}\over R_2}\delta^{ab}) 
=
{\ell_g^3\over R_2} \left ( \eta^{\alpha\beta} \oplus \lambda
 \delta^{ab}\right )\, ,
\eeq
\beq
\theta^{AB}=({\Theta_{\rm T}^{3\over 2}\over R_2}\epsilon^{\alpha\beta}
\oplus 0)
={\ell_g^3\over R_2}(\epsilon^{\alpha\beta}\oplus 0)\ ,
\eeq\beq
G_{\rm O}= {R_2 \over R_5} \sqrt{\Theta_{\rm S} \over \Theta_{\rm T}}=
{R_2\over R_5}\sqrt{\lambda}  \ ,
\eeq\beq
r_{\rm B}\equiv{R_{\rm B}^{(S)}\over \sqrt{\alpha^\prime}} =
{\Theta_{\rm S}^{{1\over 2}}
\Theta_{\rm T}^{1\over 4} \over R_2^{1\over 2} R_5} \ .\label{smc}
\eeq

Hence the NCOS limit on the D3--brane has effective open string scale
$\alpha^\prime_{\rm eff}$ and noncommutativity parameter $\theta_{\rm NCOS}$ 
both given by

\begin{equation}
\alpha^\prime_{\rm eff} \equiv \theta_{\rm NCOS} = 
{\Theta_{\rm T}^{3\over 2} \over
R_2} \, ,
\end{equation}

Importantly, the worldsheet sigma model coupling constant $r_{\rm B}$
given by (\ref{smc}) remains finite in the limit.

In the S-dual/modular transformed description we find (inserting the
five brane limit into the second set of relations):

\begin{eqnarray}
\tilde{{\cal F}}_{01} &=&0 \, , \qquad \tilde{{\cal F}}_{34} = \epsilon^{0}
\\
{ \tilde{g}_{AB} \over \tilde{\alpha}^{\prime} }&\sim&
{\rm diag}(\epsilon^{-{1\over 2}},\epsilon^{-{1\over 2}},\epsilon^{{1\over
2}},\epsilon^{{1\over 2}})
\, , \qquad \tilde{g}_s \sim \epsilon^{{1\over 2}}
\end{eqnarray}

This we identify with the noncommutative field theory limit as described
in \cite{sw}. In this case the mass scales on the D3--brane are sent
to zero, i.e.~$\tilde{\alpha}'G^{AB} \rightarrow 0$, decoupling the
massive string modes.

The relations between the fixed M--theory quantities and the fixed
quantities on the
D3--brane are as follows:

\begin{eqnarray}
g_{YM}^2&=& {R_5 \over R_2 \sqrt {\lambda}} 
\ ,\qquad \theta_{\rm NCYM} = 
{\Theta_{\rm S}^{3\over 2} \over R_5} \ ,\\
m &=& {\Theta_{\rm S}^{1\over 4} \over \sqrt{R_5} R_2} \ ,
\end{eqnarray}

where $m$ is the periodicity in mass units of the compact scalar in the
four--dimensional noncommutative action. Note that for the
noncommutative field theory, instead of a fixed worldsheet sigma model
radius $r_{\rm B}$ we now find a
fixed kinetic term for the compact transverse scalar $\Phi \equiv {X^9
\over
R_B^{(S)} } $ in the limit when $\epsilon\rightarrow 0$. 


The natural fixed moduli for the noncommutative M-five brane are $l_g$,
the complex structure of the torus, $\tau_{OM}$ and the area of the
torus, $A_{OM}$ as measured by the open membrane metric:
\beq
\tau_{OM} ={R_2 \over R_5} \sqrt{\lambda}\, , \quad A_{OM}= R_2 R_5 {1
\over
\sqrt{\lambda}}
\eeq

We now recover the standard relations between M theory and the S-dual
descriptions of IIB for the noncommutative open string/membrane moduli.
For the noncommutative open string,
\begin{eqnarray}
G_O &=& \tau_{OM} \\ \nonumber
r_B &=& A_{OM}^{-{3\over4}} \tau_{OM}^{1\over4} l_g^{3\over2} \, .
\end{eqnarray}
For the S-dual, noncommutative field theory,
\begin{eqnarray}
g^2_{YM} &=& {1 \over \tau_{OM}} \\ \nonumber
m  &=&   A_{OM}^{-{3\over4}} \tau_{OM}^{-{1\over4}}
l_g^{1\over2}\, .
\end{eqnarray}
The duality transformation is now obtained by a modular transformation on
the torus as seen by the open membrane metric so that the two theories are
related by:
\beq
\tau_{OM} \rightarrow {1 \over \tau_{OM}}
\eeq
with the appropriate interpretation of duality related quantities.
We remark that the couplings are independent of our choice of conformal
factor for the open membrane metric.

The duality between NCOS and NCYM is possible for the D3--brane because both
open string coupling and Yang-Mills coupling are independent of the closed
string scale $\alpha'$.

Finally, we consider the following limits of the NCOS on the
D3--brane (we set $\lambda=1$ below):

\begin{enumerate}
\item[1)] $T$: Taking $r_{\rm B}\rightarrow 0$ while keeping $G_{\rm
O,A}\equiv
r_{\rm B}^{-1}G_{\rm O}$ fixed leads to the NCOS on the T-dual D4--brane
with open string coupling $G_{\rm O,A}$. This is analogous to how the
usual closed string coupling transforms under T-duality. It is interesting
that this noncommutative open string theory exhibits a sort of T-duality.

\item[2)] $S$: Taking $G_{\rm O}\rightarrow \infty$ while keeping
$\theta_{\rm
NCYM}\equiv G_{\rm O}\sqrt{\alpha'_{\rm eff}}$ fixed leads to the S-dual
NCYM on the D3--brane with $g_{YM}\rightarrow 0$, as expected.

\end{enumerate}

\section{Discussion}

We have argued for the existence of a decoupled
noncommutative theory on the five--brane defined by the limit
(\ref{tnc5lim}) by showing its relation to various well--defined limits
of IIA and IIB string theory. Ultimately we are of course interested in
finding an intrinsically six--dimensional definition of the
decoupled theory. One may wonder to what extent
the open membrane action underlies such a formulation and in particular
whether there is an analog of the subtle cancellations
between the diverging electric field and tension that occur in the
string case. In the critical
limit we expect the finite parts of the Wess--Zumino term to yield the
non--commutative structure of the five--brane loop--space
via the definition
of the functional Moyal product given in \cite{us}. The role of
the kinetic part of the action
is more unclear, however, due to the usual membrane instability.
Interestingly one may construct a stable, non--degenerate open membrane
solution in the
critical limit which is {\it not} a limit of any solution for finite
$\epsilon$. This is the analogue of the the critical string solution
discussed in \cite{lennat}. One would hope that the quantisation of the
membrane in this near critical background will provide the open membrane
metric with the appropriate conformal factor given in this paper. This is
ongoing
work. It is not yet clear whether the critical field will cure the usual
membrane sicknesses.

Finally we wish to make the following observation, given our choice of
conformal factor for the open membrane metric we see that the line element
for the self--dual string solution \cite{Howe:1998ue} is exactly 
$AdS_3\times S^3$ in the near horizon limit. Given that the proceedure in
this paper has been to reproduce the usual bulk relations for the
decoupled theories on the brane one wonders whether it might be possible
to formulate an AdS/CFT corresspondence \cite{maldacena} for the
self--dual
string in the
five brane. 

In summary, the NCOS on the D4--brane
with noncommutativity parameter $\theta=\alpha'_{\rm eff}$ has a dual
description in the limit of strong coupling $G_O>>1$ as a
noncommutative five--brane with fundamental length
$\ell_g=G_O^{1\over 3}\sqrt{\alpha'_{\rm eff}}$ reduced on a circle of
radius
$R=G_O\sqrt{\alpha'_{\rm eff}}$. The S--duality of the NCOS and
NCYM theories on a directly reduced D3--brane follows from the modular   
invariance of the noncommutative five--brane wrapped on a torus. The
couplings on the D3--brane are identified with the complex structure of
the torus in the open membrane metric.

\bigskip

{\bf{Note Added}}
\bigskip

During the completion of this paper we received the preprint 
\cite{gmns} that also discusses the noncommutative open
membrane limit and its relation to the noncommutative open
string limit on the D4--brane. The preprint \cite{gmns} also contains an
interesting discussion of NCOS theories at strong coupling
for the other D--branes. Our paper emphasizes the role played
by the open membrane metric and the relation between M/IIB moduli.

\bigskip
\centerline{\bf Acknowledgements}
\bigskip

P.S.~is grateful to M.~Cederwall, H.~Larsson and B.E.W.~Nilsson for
discussions. D.S.B.~thanks R.~Gopakumar for discussions during
the Fradkin Memorial conference on {\it Quantization, Gauge Theory and
Strings}. This work is supported by the European Commission TMR program
ERBFMRX-CT96-0045, in which E.B., D.S.B.~and J.P.v.d.S.~are associated to
the University of Utrecht. The work of J.P.v.d.S.~and P.S.~is part of the
research program of the ``Stichting voor Fundamenteel Onderzoek der
Materie'' (FOM).


\begin{thebibliography}{99}

\bibitem{Connes:1998cr}
A.~Connes, M.~R.~Douglas and A.~Schwarz,
``Noncommutative geometry and matrix theory: Compactification on tori,''
JHEP {\bf 9802} (1998) 003
[hep-th/9711162];\\
M.~R.~Douglas and C.~Hull,
``D-branes and the noncommutative torus,''
JHEP {\bf 9802} (1998) 008
[hep-th/9711165];\\
Y.~E.~Cheung and M.~Krogh,
``Noncommutative geometry from 0-branes in a background B-field,''
Nucl.\ Phys.\  {\bf B528} (1998) 185
[hep-th/9803031].

\bibitem{Schwarz:1998qj}
A.~Schwarz,
``Morita equivalence and duality,''
Nucl.\ Phys.\  {\bf B534} (1998) 720
[hep-th/9805034];\\
C.~Hofman and E.~Verlinde,
``U-duality of Born-Infeld on the noncommutative two-torus,''
JHEP {\bf 9812} (1998) 010
[hep-th/9810116].

\bibitem{sw}
N.~Seiberg and E.~Witten,
``String theory and noncommutative geometry,''
JHEP {\bf 9909} (1999) 032, 
[hep-th/9908142].

\bibitem{GMSS}
R.~Gopakumar, J.~Maldacena, S.~Minwalla and A.~Strominger,
``S-duality and Noncommutative Gauge Theory'',
[hep-th/0005048].

\bibitem{lennat}
N.~Seiberg, L.~Susskind and N.~Toumbas,
``Strings in Background Electric Field, Space/Time Noncommutativity
and A New Noncritical String Theory'',
[hep-th/0005040].


\bibitem{Barbon:2000sg}
J.~L.~Barbon and E.~Rabinovici,
``Stringy fuzziness as the custodian of time-space noncommutativity,''
hep-th/0005073.


\bibitem{Chen:2000ny}
G.~Chen and Y.~Wu,
``Comments on noncommutative open string theory: V-duality and  holography,''
hep-th/0006013.


\bibitem{kleb}
I.~R.~Klebanov and J.~Maldacena, 
``1+1 Dimensional NCOS and its U(N)
Gauge Theory Dual,''
hep-th/0006085. 



\bibitem{troels}
 T.~Harmark, 
``Supergravity and space-time
non-commutative open string theory,''
[hep-th/0006023].



\bibitem{us}
E.~Bergshoeff, D.~S.~Berman, J.~P.~van~der~Schaar and P.~Sundell,
``A Noncommutative M-theory Five-brane'',
[hep-th/0005026].


\bibitem{Howe:1997mx}
P.~S.~Howe and E.~Sezgin,
``Superbranes,''
Phys.\ Lett.\  {\bf B390} (1997) 133
[hep-th/9607227];\\
P.~S.~Howe and E.~Sezgin,
``D = 11, p = 5,''
Phys.\ Lett.\  {\bf B394} (1997) 62
[hep-th/9611008];\\
P.~S.~Howe, E.~Sezgin and P.~C.~West,
``Covariant field equations of the M-theory five-brane,''
Phys.\ Lett.\  {\bf B399} (1997) 49
[hep-th/9702008].
\\
E.~Sezgin and P.~Sundell,
``Aspects of the M5-brane,''
hep-th/9902171, proceedings of the Trieste Conference on
Superfivebranes and Physics in 5+1 Dimensions, 1-3 April, 1998, Trieste,
Italy.

\bibitem{gmns}
R.~Gopakumar, S.~Minwalla, N.~Seiberg and A.~Strominger,
``OM Theory in Diverse Dimensions'',
[hep-th/0006062].
















\bibitem{schwarz}
J.~H.~Schwarz,
``An SL(2,Z) multiplet of type IIB superstrings,''
Phys.\ Lett.\  {\bf B360} (1995) 13
[hep-th/9508143].


\bibitem{aspinwall}
P.~S.~Aspinwall,
``Some relationships between dualities in string theory,''
Nucl.\ Phys.\ Proc.\ Suppl.\  {\bf 46} (1996) 30
[hep-th/9508154].


\bibitem{Berman:1998va}
D.~S.~Berman,
``M5 on a torus and the three brane"
Nucl. \ Phys. \ {\bf B533} (1998) 317
[hep-th/9804115].














\bibitem{Howe:1998ue}
P.~S.~Howe, N.~D.~Lambert and P.~C.~West,
``The self-dual string soliton,''
Nucl.\ Phys.\  {\bf B515}, 203 (1998)
[hep-th/9709014].







\bibitem{maldacena}
J.~Maldacena,
``The large N limit of superconformal field theories and supergravity,''
Adv.\ Theor.\ Math.\ Phys.\ {\bf 2} (1998) 231 [hep-th/9711200].


\end{thebibliography}
\end{document}